\documentstyle[aps,prl,floats,twocolumn,psfig,epsf]{revtex}

\begin{document}
\draft

\twocolumn[\hsize\textwidth\columnwidth\hsize\csname
  @twocolumnfalse\endcsname

\title{Theory of Tunneling Spectroscopy in Ferromagnetic Nanoparticles}

\author{C.~M.~Canali$^{1}$ and A.~H.~MacDonald$^{2}$}

\address{$^1$Division of Solid State Theory,
Department of Physics, Lund University, SE-223 62 Lund, Sweden\\
    $^2$Department of Physics, Indiana University, Bloomington,
    Indiana, 47405}

\date{\today}

\maketitle

\begin{abstract}

We present a theory of the low-energy excitations of a
ferromagnetic metal nanoparticle.  In addition to the
particle-hole excitations, which occur in a paramagnetic metal
nanoparticle, we predict a branch of excitations involving the
magnetization-orientation collective coordinate.  
Tunneling matrix elements are in general sizable for several
different collective states associated with the same band
configuration.  We point out that the average change in
ground state spin per added electron differs from non-interacting
quasiparticle expectations, and that the change in the 
spin-polarization, due to Zeeman coupling, is strongly
influenced by Coulomb blockade physics.

\end{abstract}

]

\narrowtext

The energies of many-particle states in metallic nanoparticles and
semiconductor quantum dots can be measured directly by tracking the
dependence of resonant tunneling conductance peaks on gate and bias
voltages\cite{kastner}. This technique has been used to study
the interaction physics of quantum dots at weak fields\cite{weakfields}
and in the quantum Hall regime\cite{ashoorireview}, and metallic  
nanoparticles in both superconducting\cite{ralphsc} and normal\cite{normal}
states.  In zero-field semiconductor quantum dots at high densities
and in normal metallic nanoparticles, experiments are generally consistent
with a model which acknowledges electron-electron interactions
only in a mean-field electrostatic term that, because of its
long range, gives rise to the Coulomb blockade
effect\cite{cbreviews}.  The success of this simple interpretation
is a consequence of the Fermi-liquid character of
these interacting electron systems.

The present work is motivated by recent experimental studies\cite{ralphfm} of
tunneling via discrete energy levels in ferromagnetic Cobalt
nanoparticles that find resonance spacings smaller than predicted in an
independent particle picture, and a dependence on external field 
qualitatively different from
those in paramagnetic metal nanoparticles.  Since bulk ferromagnetic
metals have low-energy spin-wave excitations in addition to their
Fermi liquid particle-hole excitations, it is natural, as
suggested\cite{ralphfm} by Gu\'eron {\it et al.}, to seek an
explanation in terms of the collective quantum physics of the
magnetic order parameter field.  In this Letter we address the
inter-play of quasiparticle and collective order parameter excitations
in tunneling spectroscopy studies of ferromagnetic nanoparticles.
Our conclusions are based in part on the properties of a simple
exactly solvable toy model, described in the following paragraphs.
We conclude that only states in which all singly occupied
nanoparticle orbitals have aligned spins are
relevant at low-energies in ferromagnetic nanoparticles, and use
this observation to derive expressions for the tunneling amplitudes
of ground and excited many-particle collective spin states.
Near fields where magnetization reversal occurs,
many of these low-energy spin states have large tunneling amplitudes, 
and contribute significantly to the tunneling spectrum, partly explaining the 
enhanced density of resonances seen in experiment.

In ferromagnetic metals, short-range exchange interactions
favor spin-alignment, giving rise to an approximately rigid spin-splitting
of quasiparticle energies in the ferromagnetic ground state.  Our
toy model reflects this rigidity by assuming identical exchange
constants between all pairs of orbitals:
\begin{equation}
{\cal H} = \sum_{j,\sigma} c^{\dagger}_{j,\sigma} c_{j,\sigma} \epsilon_j
- \frac{U}{8} \sum_{j,k} \sum_{s,s',t,t'} c^{\dagger}_{j,s'}
{\vec \tau}_{s',s} c_{j,s} \cdot c^{\dagger}_{k,t'} {\vec \tau}_{t',t}
c_{k,t}.
\label{toyhamiltonian}
\end{equation}
In Eq.~\ref{toyhamiltonian}
$\epsilon_j$ is a nanoparticle orbital energy that incorporates the
charge correlation physics neglected in our model Hamiltonian and
${\vec \tau}$ is the Pauli spin matrix vector.  The single
particle orbitals have an average spacing inversely
proportional to the volume of the nanoparticle and are expected to
exhibit spectral rigidity\cite{levelstats}.  The many-particle
spectrum of this Hamiltonian follows readily from the following
observations: i) the total occupation of each orbital is a 
good quantum number ii) the
interaction term is proportional to the square of the total
electron spin operator $\vec S_{tot}$. The $2^{N_s}$ states with a
given set of $N_s$ singly occupied orbitals have their band energy
degeneracy lifted by the interaction energy $- (U/2) S_{tot}
(S_{tot}+1)$, where the total spin $S_{tot}$ has a maximum value
$N_s/2$. We show below that only this $N_s+1$-fold degenerate
spin-multiplet is relevant to the low energy physics of
a ferromagnetic nanoparticle.

We start by considering the ground state of a ferromagnetic
metal nanoparticle in which majority and minority spin
quasiparticles are occupied up to their Fermi energies. 
We take these to have the values 
$\epsilon_{Fa}$ and $\epsilon_{Fi}$ in the absence of an external
magnetic field at a reference total particle number.
We are interested
in how the nanoparticle evolves as a function of external field
and gate voltage.  We assume, for simplicity, that the
quasiparticle energy levels of majority and minority spins are
equally spaced near their respective Fermi energies.
It follows from the considerations of the
previous paragraph that the total energy, relative to that of
the reference state, is
\begin{eqnarray}
\delta E &=& [\epsilon_{Fa} - \Delta/2 + \delta_{a}/2] \delta N_{a}
         + [\epsilon_{Fi} + \Delta/2 + \delta_{i}/2] \delta N_{i}
         \nonumber \\
         &+&E_{cb} (\delta N_{a} + \delta N_{i})^2/2 \nonumber \\
         &+& \frac{\delta N_{a}^2}{2} (\delta_{a} - \Delta/2(2S_0+1))
         + \frac{\delta N_{i}^2}{2} (\delta_{i} - \Delta/2(2S_0+1))
         \nonumber \\
         &+& \delta N_{a} \delta N_{i}  \frac{\Delta}{2(2S_0+1)}\nonumber \\
         &-& (V_g  + g \mu_B B/2) \delta N_{a} - (V_g - g \mu_B B/2)
         \delta N_{i}.
\label{grainenergy}
\end{eqnarray}
In this energy expression, $\delta_{a,i}$ are level spacings, and
we have added by hand the electrostatic
Coulomb blockade term, which depends only on the total number of
particles in the grain.
The terms proportional to $\Delta$ in Eq.~\ref{grainenergy}
originate from the interaction energy $-U S_{tot}(S_{tot}+1)/2$;
the notational change $U \to \Delta/(2 S_0+1)$ is motivated by the
identification, explained below, of $\Delta$ as the spin-splitting energy of the
quasiparticle bands.  Here $S_0 = (N_{a0}-N_{i0})/2$ is the ground
state total spin of the reference state and $S_{tot}=S_0 + (\delta
N_{a} - \delta N_{i})/2$.  Note that for a ferromagnetic particle $S_0$
will be proportional to volume.  We have not
explicitly indicated the capacitance ratio relating the gate
voltage and the chemical potential of the nanoparticle, and have
assumed that only Zeeman coupling to an external field is
important. 

Stability of the reference
system ground state requires that its energy
increases at fixed total particle number $N$ both for $S_{tot} \to S_0 +1$
($\delta N_{a} =1, \delta N_{i}=-1$) and $S_{tot} \to S_0 -1$
($\delta N_{a}=-1, \delta N_{i}=+1$).  From this it follows that
\begin{equation}
| \epsilon_{Fa} - \epsilon_{Fi} - \Delta +
(\delta_{a}-\delta_{i})/2 | < \big[\delta_{a}+\delta_{i} - 2 \Delta/(2
S_0 +1)\big].
\label{stability}
\end{equation}
Since the right hand side of Eq.~\ref{stability} $\sim N_A^{-1}$,
where $N_A$ is the number of atoms in the grain, it follows that
$\Delta = \epsilon_{Fa} - \epsilon_{Fi}$ to within a fluctuating
mesoscopic correction, {\it i.e.} that $\Delta$ is the quasiparticle
spin-splitting.
The energy difference between the $S_{tot} = (N_{a}-N_{i})/2$
states retained in our considerations and the $S_{tot} \le (N_{a}-N_{i})/2 -1$
states we have discarded is $\Delta 2 S_0/ (2S_0+1) \sim \Delta$,
well outside the energy range of interest\cite{spinwave}.

It is instructive to minimize the energy by considering $N_{a}$
and $N_{i}$ as continuous variables, thereby obtaining the trend
lines around which the mesoscopic ground-state
spin and charge quantum numbers fluctuate as gate voltage
and external field vary.
We find that
\begin{eqnarray}
C^{-1}_{aa} \delta N_{a} + C^{-1}_{ai} \delta N_{i} &=& V_{g} + g\mu_B B
\nonumber \\
C^{-1}_{ia} \delta N_{a} + C^{-1}_{ii} \delta N_{i} &=& V_{g} - g\mu_B B
\label{trends}
\end{eqnarray}
where
\begin{eqnarray}
C^{-1}_{aa} &=& E_{cb} + \delta_{a} - \Delta/4S_0 \nonumber \\
C^{-1}_{ai} &=& C^{-1}_{ia} = E_{cb} + \Delta/4S_0 \nonumber \\
C^{-1}_{ii} &=& E_{cb} + \delta_{i} - \Delta/4S_0.
\label{inversecapacitance}
\end{eqnarray}
The inverse capacitance matrix $C^{-1}$ describes the
coupled spin and charge response of the nanoparticle.  This coupling
is responsible for the variation
of total particle number with external field observed\cite{ono,sitges}
by Ono {\it et al.}.

Eq.~\ref{trends} can be solved to obtain the
variation of both $N_{a}$ and $N_{i}$ with either $V_g$ or $B$.
For example, we find that
\begin{equation}
\frac{dS_{tot}}{d(g\mu_BB)} = \frac{2 E_{cb} +
(\delta_{a}+\delta_{i})/2}{E_{cb}(\delta_{a}+\delta_{i} - \Delta/S_0)
+ \delta_{a} \delta_{i} - \Delta (\delta_{a}+\delta_{i})/4S_0},
\label{sovsb}
\end{equation}
and that
\begin{equation}
\frac{dN_{a}/dV_g}{dN_{i}/dV_g} =
\frac{M_{ii}-M_{ai}}{M_{aa}-M_{ai}}= \frac{\delta_{i} -
\Delta/2S_0}{\delta_{a} - \Delta/2S_0}.
\label{nadnivsvg}
\end{equation}

Level spacings and exchange splittings in metallic ferromagnet nanoparticles
can be estimated from spin-density functional band structure
calculations.
For the case of Cobalt particles\cite{papabook}
$\delta_{a}= 5.55 {\rm eV}/N_A$, $\delta_{i}= 1.43 {\rm eV}/N_A$,
$2S_0 = 1.65 N_A$, and $\Delta/2S_0 = 1.07 {\rm eV}/N_A$.  The Coulomb blockade
energy is sensitive to the screening environment of the
nanoparticle; for a grain with $N_A \approx 1500$, Ralph {\it et
al.} find that $E_{cb} \ge 30 {\rm meV}$.  As expected, the largest
elements in the inverse capacitance matrix are the purely
electrostatic Coulomb blockade contributions.  Evaluating the right hand side
of Eq.~\ref{nadnivsvg} we find that $S_{tot}$ almost always {\em decreases}
when particles are added. 
The response of $S_{tot}$ to external fields, given by
Eq.~\ref{sovsb}, is smaller than would be expected naively,
because $E_{cb}$, the dominant mesoscopic energy scale, suppresses
charge fluctuations causing the larger (majority spin) level
spacing to limit this response.  Typically the relatively large value of
$E_{cb}$ suppresses the total charge response to both gate voltage and
external field.  For example, for $N_A=1500$ and $E_{cb}=30{\rm meV}$,
typical for the particle size studied by Gu\'eron {\it et al.},  we find
that $\delta N_{a} = 0.0025 d V_{g} + 0.306 d (g \mu_B B)$ and $\delta N_{i}
= 0.0302 d V_{g} - 0.321 d (g \mu_B B)$ with energies in meV.

In the model discussed so far, structure would appear in the 
bias dependence of the tunneling current only on the scale of the 
single-particle level spacing.  Following Gu\'eron {\it et al.},
we seek an explanation of experiment by invoking 
spin-orbit coupling and long-range dipole interactions
between the electronic spins.  
%The many-particle
%Hamiltonian is then not spin-rotational invariance and the
%many-particle spectrum no-longer consists of spin-multiplets with
%macroscopic degeneracies.
In Cobalt, the typical band energy shift due to spin-orbit coupling is\cite{koelling} 
$\delta E \sim 1 {\rm meV}$; the dependence of the total band energy
on spin-moment orientation results from a partially cancelling  
sum of spin-orbit energy shifts over all singly occupied orbitals.
The typical spin-orbit matrix element between an individual pair of orbitals
in a grain is $\sim \delta E/N_A$, $ \sim 1 {\rm \mu eV}$ for the grain sizes
of interest to us.  These considerations justify starting from a model which 
neglects spin-orbit coupling of states with different band configurations 
and uses the bulk magnetocrystalline anisotropy coefficient to 
describe its effect within a single multiplet:
\begin{equation}
{\cal H} = {\cal H} - K (S_z/S_0)^2.
\end{equation}
where, for Cobalt\cite{coey}, $K \sim 0.073 N_A\,{\rm meV}$.
% AHMd4: This comment can be dropped if required for space.
Note for $N_A \sim 1500$, $K \sim .1 {\rm eV}$, much larger than either
level spacing or Coulomb blockade energy scales.

\begin{figure}[tbp]
\begin{center}
\hspace{0.2truecm}
\psfig{file=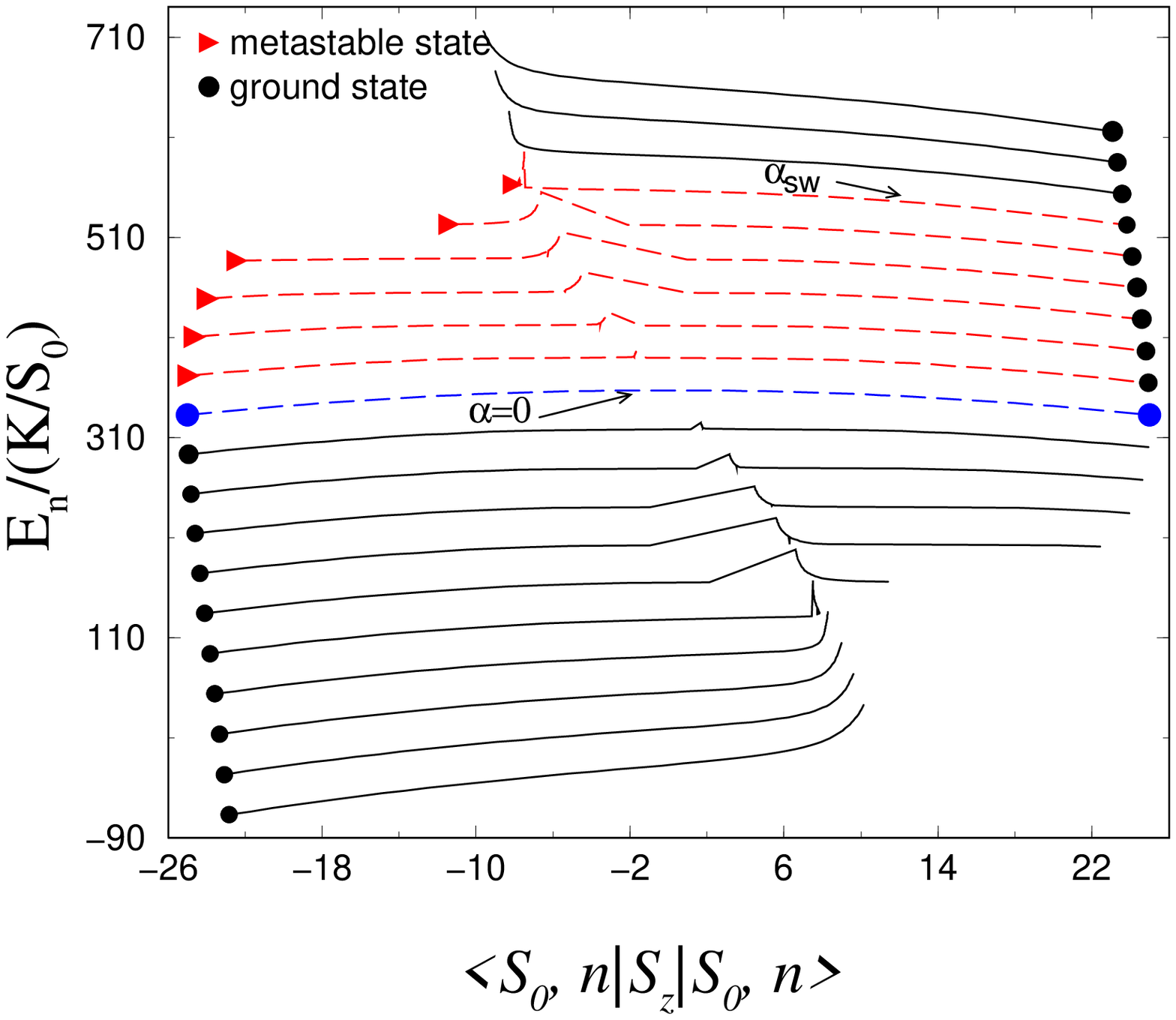,height=2.6in,width=2.6in}
\end{center}
 \caption{Energy eigenvalues $E_n$ vs. $\langle S_0, n|S_z|S_0, n\rangle$ for
the collective spin Hamiltonian. The curves, offset for clarity,
correspond to increasing values of $\alpha$, starting from $\alpha=-2$
at the bottom.
Here $S_0 =25$ and $\hat B$
is oriented at $\pi/4$ from the easy axis, $\hat z$. 
At $\alpha= \alpha_{\rm sw}\sim 1.2$, the metastable state
disappears.}
\label{fig:one}
\end{figure}

A theory of tunneling spectroscopy in ferromagnetic nanoparticles
requires results for tunneling matrix elements between many-body 
eigenstates,
found by diagonalizing the collective spin Hamiltonian
\begin{eqnarray}
{\cal H} =&& E_{\rm band} - K (S_z/S_0)^2 - g \mu_B {\vec B}\cdot {\vec S}\\
=&& E_{\rm band} -K/S_0\Big( S_z^2/S_0 + \alpha \hat B\cdot {\vec S}\Big)
\label{colspinham}
\end{eqnarray}
where $K/S_0\approx 0.05{\rm meV}$ is the natural energy scale
and $\alpha \equiv g \mu_B B/(K/S_0)\sim 2\,B[{\rm T}]$ is
the (dimensionless) strength of the magnetic field.
In Fig.~\ref{fig:one}
we plot the energy eigenvalues $E_n$ as a function of the
expectation value of $S_z$ %   and $S_B/S_0 \equiv  \vec S \cdot \hat B/S_0$
for the corresponding eigenstates $|S_0, n\rangle$ for $S_0=25$ and $\hat B$
oriented at $\theta_{\rm ext}=\pi/4$ from the easy axis, $\hat z$.  
The curves are offset for clarity, starting at the 
bottom with $\alpha =-2$. For large negative $\alpha$
the ground state spin is polarized along the field direction, 
and gradually reorients toward $-\hat z$ as the field is ramped to zero.
At $\alpha=0$ a level crossing occurs and the ground state state 
is now polarized along $\hat z$.
The former ground state, now classically metastable and separated  
from the true ground state by a potential barrier, is apparent in the 
spectrum until the classical switching field is reached at 
$\alpha_{SW} \sim +1.2$.  The classical down-sweep metastable states 
appear at negative values of $\alpha$ and positive values of 
$\langle S_0, n | S_z |S_0, n \rangle$.

In order to evaluate the many-particle tunneling matrix elements,
%it is necessary to express these eigenstates in terms of
we express these eigenstates in terms of
microscopic electronic degrees of freedom.  
For a given set of
quasiparticle occupations, the microscopic state for collective
spins oriented in direction $\hat \Omega= \hat \Omega(\theta,\phi)$ is
\begin{equation}
|\Psi(\hat \Omega)\rangle = \prod_{j \in S} [{\cal V}_{\uparrow}(\hat \Omega)
c^{\dagger}_{j,\uparrow} + {\cal V}_{\downarrow}(\hat \Omega)
c^{\dagger}_{j,\downarrow}] \prod_{k \in D}
c^{\dagger}_{k,\uparrow} c^{\dagger}_{k,\downarrow} |0\rangle,
\label{manyelectronstate}
\end{equation}
where $D$ and $S$ are the sets of doubly and singly occupied orbitals 
respectively; 
%The functions $\cal V_{\uparrow}$ and $\cal V_{\downarrow}$ 
%are defined as
${\cal V}_{\uparrow}(\hat \Omega) = \cos(\theta/2)\exp(i\phi)$
and ${\cal V}_{\downarrow}(\hat \Omega)=
\sin(\theta/2)\exp(-i\phi)$.

The matrix element between states with collective spin orientations
$\hat \Omega'$ and $\hat \Omega$ for adding an electron with spin-orientation
$\hat \Omega_{\rm ext}$ to an empty orbital, 
thereby increasing $N_s$ and $S_{tot}$, is
\begin{equation}
\langle \Psi(\hat \Omega')|c^{\dagger}(\hat \Omega_{\rm ext})|
\Psi(\hat \Omega) \rangle_{+} = \langle \hat \Omega' | \hat
\Omega\rangle_{1/2}^{N_s} \langle \hat \Omega' |\hat
\Omega_{\rm ext}\rangle_{1/2}\; ,
\label{meplus}
\end{equation}
where $\langle \hat \Omega' | \hat\Omega\rangle_{1/2}$ is the
inner product of spin-1/2 coherent states\cite{auerbach}.
\begin{figure}[tbp]
\begin{center}
\hspace{0.2truecm}
\psfig{file=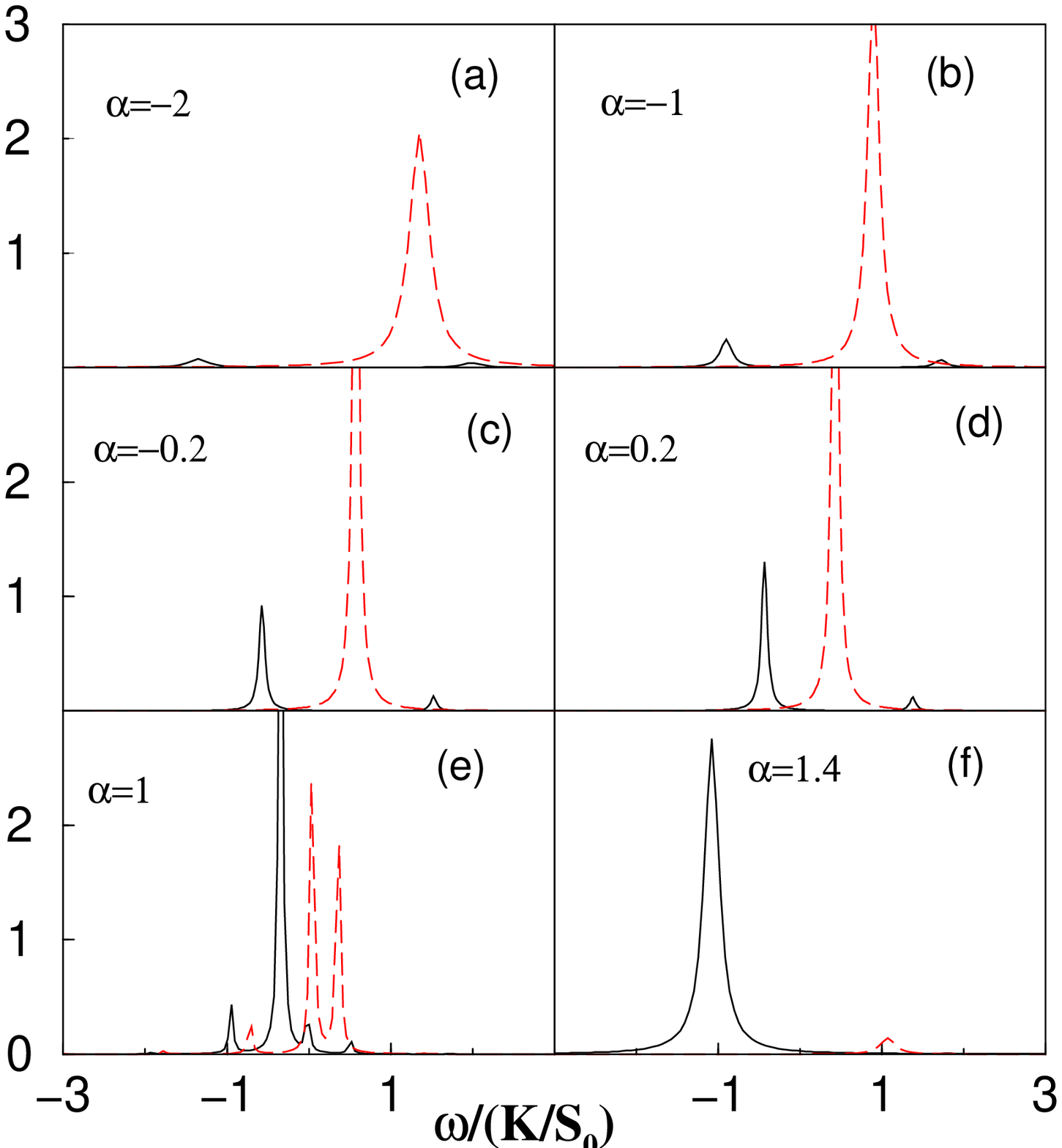,height=2.6in,width=2.6in}
\end{center}
 \caption{The tunneling spectral function for the collective spin model,
at different values of the magnetic field, specified by $\alpha$.
The solid and dashed lines refer to the $S_{\rm tot} \to S_0\pm 1$ transitions
respectively.
$S_0 =25$ and $\hat B$
is oriented at $\pi/4$ from the easy axis $\hat z$.
For $0< \alpha < \alpha_{\rm sw} \sim 1.2$, 
the system is in a metastable state (see Fig.~\ref{fig:one}),
before the tunneling takes place. In (e), $\alpha$ is near $\alpha_{\rm sw}$,
where the metastable state disappears and the magnetization reversal occurs.}
\label{fig:two}
\end{figure}
Similarly the matrix element for adding an electron with
spin-orientation $- \hat \Omega_{\rm ext}$ to a singly occupied orbital,
thereby decreasing $N_s$ and $S_{tot}$ is
\begin{equation}
\langle \Psi(\hat \Omega^{'})|c^{\dagger}(\hat \Omega_{\rm ext})| \hat
\Psi(\hat\Omega) \rangle_{-} = \langle \hat \Omega' | \hat
\Omega\rangle_{1/2}^{N_s-1} \langle \hat \Omega |\hat
\Omega_{\rm ext}\rangle_{1/2}.
\label{meminus}
\end{equation}

The tunneling spectral function is defined as
\begin{eqnarray}
A_{\pm}(\omega) = &&\sum_n\Big|\langle {(N_s \pm 1)/2, n}|
c^{\dagger}(\hat\Omega_{\rm ext})|{N_s/2, 0}\rangle\Big|^2\nonumber \\
&& \times\delta\Big[\omega - (E_{(N_s \pm 1)/2, n} -E_{N_s/2,0})\Big]\; ,
\label{spectralf}
\end{eqnarray}
where $ |(N_s \pm 1)/2,n\rangle$, $E_{(N_s \pm 1)/2, n}$ are collective spin
eigenstates and eigenvalues of $\cal H$ 
in the $S_{\rm tot} = S_0 \pm 1/2= (N_s \pm 1)/2$ 
manifold respectively. 
To evaluate the matrix
elements in Eq.~\ref{spectralf}, 
we expand these eigenstates first in terms of
eigenstates of $S_z$ and then in terms of spin-coherent
states\cite{auerbach} with a definite $\hat \Omega$. 
Using Eqs.~\ref{meplus},\ \ref{meminus}, 
%and the explicit expression 
%for the inner product
%of spin-coherent states\cite{auerbach}, 
we finally arrive at the simplified expressions
\begin{eqnarray}
&&\langle {(N_s \pm 1)/2, n}|
c^{\dagger}(\hat\Omega_{\rm ext})|{N_s/2, 0}\rangle=
\sum _{\sigma}\sum_{m'}\sum_{m}{\cal V}_{\sigma}(\hat \Omega_{\rm ext})\nonumber \\
&&\times\, c^{m'}_n(N_s+1)c^m_0(N_s)
\Big \langle {N_s \pm 1\over 2}\, m'|{N_s\over 2}\,m\, ; 
{1\over2}\,\sigma\Big \rangle\; ,
\label{spectralfCG}
\end{eqnarray}
where $c^m_n(N_s)$ are the expansion coefficients of $|N_s /2,n\rangle$;
$\Big \langle {N_s \pm 1\over 2}\, m'
|{N_s\over 2}\,m\, ;{\,1\over2}\,\sigma\Big \rangle$
are Clebsch-Gordan coefficients for adding the angular momenta
$S_0 = {N_s/ 2}$ and 
$s=1/2$.

Many-particle
tunneling spectra for $S_0 =N_s/2= 25$ and $\theta_{\rm ext} = \pi/4$ 
are illustrated in Fig.~\ref{fig:two}:
In these `up-sweep' calculations, we chose $|{N_s/ 2},0\rangle$ to be the true
ground state for $\alpha<0$ and $\alpha > \alpha_{\rm sw}$. In the
field interval $0<\alpha<\alpha_{\rm sw}$, however, 
we allowed the system to be initially
in the metastable stable, discussed in Fig.~\ref{fig:one}.
These figures clearly show that the tunneling spectra 
have a regime near the classical switching field,
where many excited states contribute
giving rise to a dense tunneling spectrum, like that seen in experiment.

For a particular ferromagnetic metal particle at a given 
gate voltage, the onset tunneling spectra can be dominated by
an intermediate state with either increased or decreased particle
number.  In either case, the lowest energy multiplet %\cite{forbidden}
could have $S_{\rm tot}=S_0 \pm 1/2$,  although decreases will be more
common for particle addition and increases for particle removal,
as we have discussed.  The calculations presented here can 
explain a dense tunneling spectra with level spacing 
$\sim K/S_0$ rather than $\delta$ near the classical switching field, 
but not at general fields.
The magnetic anisotropy energy  
landscape of a realistic nanoparticle is certain to be more
complex than in our model calculation. There is clearly
a shape anisotropy, since the particles used in the experiment 
are approximately hemispherical.
A self-consistent calculation\cite{shapeanisotropy} 
on small, irregularly shaped clusters
predicts a magnetic anisotropy energy that can be much
larger than the one we have used here. If this were the case, this energy
scale would no longer be a candidate for explaining the experiment. 
Note however that for ellipsoidal particles,
the shape anisotropy constant is comparable to
the bulk constant\cite{coey}. On the other hand,
magnetostatic interactions with nearby 
particles and more subtle surface effects\cite{finitesize}
are probably important 
but much harder to estimate. 
Finally, nonequilibrium effects\cite{agam,ralphfm} and
spin-orbit coupling between different band multiplets
could also contribute significantly to the ubiquity 
of dense tunneling spectra observed in experiment.  
%We note that the $K/S_0$ and $\delta$ energy scales are clearly but not 
%widely separated in the samples studied experimentally, 
%motivating future experiments on still smaller samples.

We are grateful to M. Deshmukh, S. Gu\'eron, D. Ralph and J. von Delft for 
helpful interactions. CMC thanks the Centre for 
Advanced Study in Oslo, Norway, for hospitality.
AHM was supported by the National Science Foundation under grant
DMR-9714055.

%\bibliographystyle{/usr/local/TeX/texmf/tex/latex/misc/prsty}
%\bibliography{/home/lulu/kjordan/biblio/bibabbrevs,/home/lulu/kjordan/biblio/bibmaster}

\begin{thebibliography}{1}
\bibitem{spinwave}
In the bulk, the toy model we are using has a collective
spin-wave spectrum with a gap $\sim \Delta$.  In a 
more realistic model the spin-wave spectrum would be
gapless without spin-orbit interactions.  In such a model,
the energy scale on which $S < N_s/2$ states would become 
important equals the spin-wave energy at a wavelength comparable
to the particle dimension.  However, this much smaller scale
is still safely outside the range of interest to us.
\end{thebibliography}

\end{document}